\pdfoutput=1
\RequirePackage{ifpdf}
\ifpdf
\documentclass[pdftex]{sigma}
\else
\documentclass{sigma}
\fi

\usepackage{xspace}

\newcommand{\eg}{e.g.,\ }

\newcommand{\NEWAXIOM}[3]{%
  \begin{description}\itemsep=0pt
  \label{#2}%
  \item[\underline{#1}]{#3}
  \end{description}
}

\newcommand{\MATH}[1]{\ensuremath{#1}\xspace}
\newcommand{\Tuple}[1]{\MATH{\left\langle{#1}\right\rangle}}
\newcommand{\EMPH}[1]{\textbf{#1}}

\newcommand{\Abs}[1]{\MATH{\left|{#1}\right|}}
\newcommand{\Norm}[1]{\MATH{\left\|{#1}\right\|}}
\newcommand{\NormM}[1]{\MATH{\Abs{#1}_\mu}}
\newcommand{\NormT}[1]{\MATH{\Abs{#1}_\tau}}

\newcommand{\MATHBB}[1]{\MATH{\mathbb{#1}}}

\newcommand{\Rset}{\MATHBB{R}}

\newcommand{\MATHIT}[1]{\MATH{\mathit{#1}}}
\newcommand{\sqspeed}{\MATHIT{v}}

\newcommand{\MATHSF}[1]{\MATH{\mathsf{#1}}}

\newcommand{\coll}{\MATHSF{coll}}
\newcommand{\inc}{\MATHSF{in}}

\newcommand{\fva}{\MATHSF{A}}  
\newcommand{\fvp}{\MATHSF{P}}  

\newcommand{\inecoll}{\MATHSF{inecoll}}

\newcommand{\out}{\MATHSF{out}}

\newcommand{\taxis}{\MATHSF{t\text{-}axis}}

\newcommand{\MATHBF}[1]{\MATH{\mathbf{#1}}}

\newcommand{\vel}{\MATHBF{v}}
\newcommand{\vp}{\MATHBF{p}}

\newcommand{\BAR}[1]{\MATH{\bar{#1}}}

\newcommand{\vx}{\BAR{x}}
\newcommand{\vy}{\BAR{y}}

\newcommand{\vu}{\BAR{u}}

\newcommand{\leteq}{\MATH{\mathrel{:=}}}
\newcommand{\gyok}{\MATH{\sqrt{\phantom{n}}}}

\newcommand{\defiff}{\MATH{\stackrel{\rm def}{\Longleftrightarrow}}}
\newcommand{\lland}{\;\land\;}

\newcommand{\setclose}{\}}

\newcommand{\setmid}{\mathrel{:}}
\newcommand{\setopen}{\{}

\newcommand{\bv}{\BAR{v}}
\newcommand{\bu}{\BAR{u}}

\newcommand{\Ip}{\MATHSF{Ip}}

\newcommand{\IOb}{\MATHSF{IOb}} 

\newcommand{\B}{\MATHIT{B}} 
\newcommand{\Ph}{\MATHSF{Ph}} 
\newcommand{\Q}{\MATHIT{Q}} 
\newcommand{\ax}[1]{\textcolor{axcolor}{\MATHSF{#1}}}
\newcommand{\li}{{\rm line}}
\newcommand{\wl}{\MATH{\mathit{w}\ell}}

\newcommand{\w}{\MATHBF{w}}

\definecolor{firebrick}{rgb}{0.7,0.13,0.13}

\definecolor{thmcolor}{rgb}{0,0,.4} 
\definecolor{remarkcolor}{rgb}{0,.2,0} 
\definecolor{proofcolor}{rgb}{.4,0,0} 
\definecolor{quecolor}{rgb}{.2,.2,0} 
\definecolor{axcolor}{rgb}{.3,0,.3}
\definecolor{thmbgcolor}{rgb}{0.9,0.9,1} 
\definecolor{rmbgcolor}{rgb}{0.9,1,0.9} 
\definecolor{proofbgcolor}{rgb}{1,0.9,0.9}

\definecolor{qcolor}{rgb}{0,0.4,0}
\definecolor{lqcolor}{rgb}{0,0.6,0}
\definecolor{phcolor}{rgb}{.6,0,0}
\definecolor{lphcolor}{rgb}{.7,0,0}
\definecolor{evcolor}{rgb}{.5,.3,.2}
\definecolor{obcolor}{rgb}{0,0.4,.5}
\definecolor{lobcolor}{rgb}{0,0.6,.75}
\definecolor{iobcolor}{rgb}{0,0,.6}
\definecolor{liobcolor}{rgb}{0,0,.7}
\definecolor{axbgcolor}{rgb}{1,.7,1}

\begin{document}

\allowdisplaybreaks

\renewcommand{\PaperNumber}{005}

\FirstPageHeading

\ShortArticleName{Relativistic Mass and Momenta of FTL Particles}

\ArticleName{Why Do the Relativistic Masses and Momenta\\ of Faster-than-Light Particles Decrease\\ as their Speeds Increase?}

\Author{Judit X.~MADAR\'ASZ~$^\dag$, Mike STANNETT~$^\ddag$ and Gergely SZ\'EKELY~$^\dag$}

\AuthorNameForHeading{J.X.~Madar\'asz, M.~Stannett and G.~Sz\'ekely}

\Address{$^\dag$~Alfr\'ed R\'enyi Institute of Mathematics, Hungarian Academy of Sciences,\\
\hphantom{$^\dag$}~P.O.~Box 127, Budapest 1364, Hungary}
\EmailD{\href{mailto:madarasz.judit@renyi.mta.hu}{madarasz.judit@renyi.mta.hu},
    \href{mailto:szekely.gergely@renyi.mta.hu}{szekely.gergely@renyi.mta.hu}}

\Address{$^\ddag$~University of Sheffield, Department of Computer Science,\\
\hphantom{$^\ddag$}~211 Portobello, Sheffield S1 4DP, United Kingdom}
\EmailD{\href{mailto:m.stannett@sheffield.ac.uk}{m.stannett@sheffield.ac.uk}}

\ArticleDates{Received September 17, 2013, in f\/inal form January 07, 2014; Published online January 11, 2014}

\Abstract{It has recently been shown within a formal axiomatic framework using a def\/i\-ni\-tion of four-momentum based on the St\"uckelberg--Feynman--Sudarshan--Recami ``switching principle'' that Einstein's relativistic dynamics is logically consistent with the existence of interacting faster-than-light inertial particles. Our results here show, using only basic natural assumptions on dynamics, that this def\/inition is the \textit{only} possible way to get a~consistent theory of such particles moving within the geometry of Minkowskian spacetime.
We present a strictly formal proof from a streamlined axiom system that given any slow or fast inertial particle, all inertial observers agree on the value of $\mathsf{m}\cdot \sqrt{|1-v^2|}$, where $\mathsf{m}$ is the particle's relativistic mass and~$v$ its speed. This conf\/irms formally the widely held belief that the relativistic mass and momentum of a positive-mass faster-than-light particle must decrease as its speed increases.}


\Classification{70A05; 03B30; 83A05}

\section{Introduction} \label{sec:introduction}

Following the introduction of Einstein's special theory of relativity in 1905, it was
generally believed that the existence of faster-than-light (FTL) particles
would violate causality, and they could not therefore exist.  Tolman's 1917 `anti-telephone'
scenario \cite{tolman} was followed by a series of papers describing causality violations involving FTL particles, leading over the last half-century to a host of related paradoxes and possible resolutions \cite{Arntzenius,BiDeSu62,ChaSil,Feinberg,Geroch,JW12,Nikolic,Pe13,Rec86,recami-ftl,Rec09,RZD04,Su70,ZRB10,ZRB12}.

\begin{figure}[t] \centering
\includegraphics{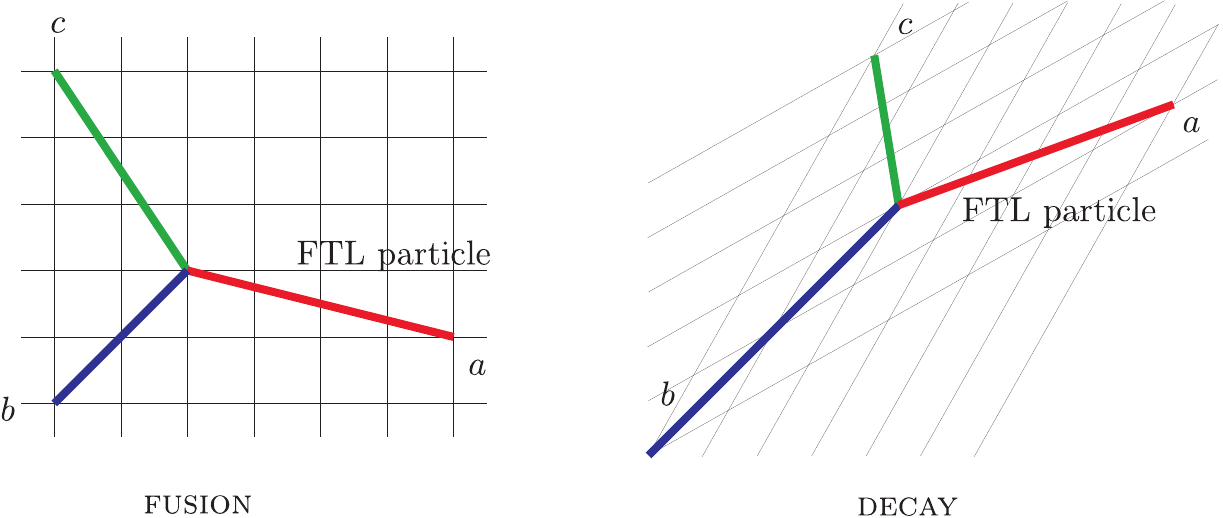}
\caption{If FTL particles are involved, what one person sees as a decay, another may see as fusion.}
\label{decay}
\end{figure}

Another issue associated with FTL particles concerns the nature of
collisions.  If one observer sees two particles (one of which is FTL) fusing
to form a third, then another, fast enough but nonetheless
slower-than-light, observer will see this fusion as the decay of one
particle into two others (see
Fig.~\ref{decay}) \cite{Rec86, Su70}.  Such ambiguities are not in themselves
paradoxical; rather, they provide one more conf\/irmation of the need
to distinguish between \textit{physical laws} and their \textit{visible
manifestations} in relativity theory \cite{BiDeSu62}.

\looseness=-1
Various authors have investigated how the relativistic mass of FTL particles ought to vary with speed, prompted in part by experimental evidence conf\/irming the existence of FTL behaviours, e.g., in certain astronomical and quantum-mechanical situations (see~\cite{Rec86} for a comprehensive review). Sudarshan~\cite{Su70} considered the theory of tachyons, and noted -- contrary to earlier assumptions in the literature~-- that their existence does not contradict any of the physical consequences of relativity theory, since these are based not on the `limiting' nature of the speed of light, but its \emph{invariance}. Even more strikingly, Recami \cite{Rec08} notes that ``at least four dif\/ferent experimental sectors of physics seem to indicate the actual existence of Superluminal motions'', two of which have been conf\/irmed both theoretically and experimentally \cite{Rec03}. In particular, \mbox{various} experiments~\cite{LLBR02, NH97,RFPM93,SKC93} (cf.~\cite{Olk04}) have conf\/irmed the quantum-theoretical prediction~\cite{AER02} that the total time taken for a photon to `tunnel' through an opaque barrier is independent of tunnel width, whence group velocities are necessarily superluminal for wide enough barriers.

The derivations presented in these earlier papers dif\/fer from ours in certain respects. For example, Recami and his colleagues follow Sudarshan \cite{Su70} in assuming that positive-energy objects travelling backwards in time cannot exist, and that ``any negative-energy particle $P$ travelling backwards in time can and must be described as its antiparticle~$\overline{P}$, endowed with positive energy and motion forward in time'', whereas we make no such blanket assumption. Recami~\cite{recami-ftl} suggests that since we know nothing a priori about tachyons, ``the safest way to formulate a theory for them is to try to generalize the ordinary theories \dots\ by performing modif\/ications as small as possible''. Our approach is to remove unnecessary assumptions so as to generalise the standard theory to the point where tachyon dynamics can be investigated without further modif\/ication.

Our derivation is, accordingly, by strict formal proof from a sparse set of basic, simple, axioms that are already known to be consistent both with standard descriptions of special relativity and with the existence of FTL particles~\cite{FTLDyn}.  We show, moreover, how the same deductive reasoning can be applied to Newtonian as well as relativistic dynamics, by replacing just one of the underlying axioms.  Our main theorem (Theorem~\ref{mainthm}) shows that
\begin{itemize}\itemsep=0pt
\item
  if relativistic dynamics holds, then all inertial observers agree on the
  value they calculate for $\mathsf{m} \cdot \sqrt{\Abs{ 1-v^2 }}$, where $\mathsf{m}$ is
  the relativistic mass of an inertial particle and $v$ is its speed, whether or not this is FTL\footnote{For `slow' particles, this invariant is the usual rest mass $\mathsf{m}_0 = \mathsf{m} \cdot \sqrt{1-v^2}$. For FTL particles, our formula $\mathsf{m} \cdot \sqrt{v^2 - 1}$ agrees with that required by non-restricted special relativity (NRR)~\cite{Rec86} (although, of course, we are working here not in NRR, but in a general axiomatic framework of dynamics).};
\item
  if Newtonian dynamics holds, all observers agree on the particle's
  relativistic mass~$\mathsf{m}$.
\end{itemize}

A consequence of this result is conf\/irmation that if an FTL inertial particle has positive relativistic mass,
its relativistic mass and momentum must decrease as its speed increases, while for slow
particles we get back the usual mass increase theorem in the relativistic
case.

In \cite{FTLDyn} a model of relativistic dynamics is constructed in which there are interacting FTL inertial particles, by extending the def\/inition of four-momentum for FTL particles using the St\"uckelberg--Feynman--Sudarshan--Recami ``switching principle'' \cite{BiDeSu62, Fey49,Stu41}. Our results here therefore complement \cite{FTLDyn}, because they show~-- using only basic natural assumptions on dynamics~-- that this def\/inition is the \textit{only} possible way to get a consistent theory of interacting FTL inertial particles.

\section{Informal formulation of our axioms and the main result}
\label{informal-formulation}

We begin with an informal outline of our basic f\/irst-order axiom system,
\ax{BA}.  This formalises the axioms used in standard approaches to special
relativity theory from Einstein onwards.  Informally paraphrased, these axioms are:
\begin{itemize}\itemsep=0pt
\item
  \ax{AxEField}\\
	Physical quantities satisfy certain basic algebraic properties of
	the real numbers.
\item
  \ax{AxL}\\
	The world-line of every inertial particle and inertial observer is a
	subset of a straight line, and contains at least two points.
\item
  \ax{AxW}\\
	The world-view transformation between any two observers is an af\/f\/ine
	transformation which takes the world-line of each body according to
	the f\/irst observer to the world-line of the same body according to
	the second observer.
\item
  \ax{AxSelf^-}\\
	Inertial observers are stationary (they do not move spatially) in their own coordinate systems.
\end{itemize}

Using \ax{BA}, we can def\/ine the speeds and velocities of inertial observers
and inertial particles according to any observer, as well as the linear momenta of inertial particles moving at f\/inite relative speeds.

\begin{figure}[t] \centering
\includegraphics{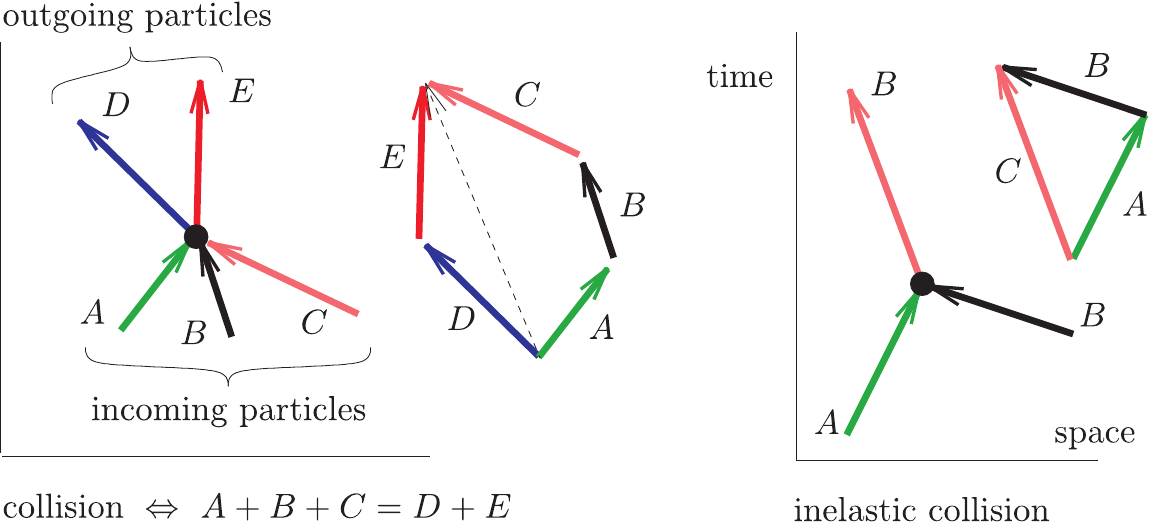}
\caption{Collision and inelastic collision of incoming and outgoing particles.  Four-momentum is conserved in collisions.}
\label{ppcoll1}
\end{figure}

Our next axiom system, \ax{DYN} (for dynamics), adds concepts relating to
particle collisions.  Intuitively, by a \textit{collision} we mean
a set of incoming and outgoing inertial particles for which the sums of the
relativistic masses and linear momenta of the incoming particles coincide
with those of the outgoing ones (Fig.~\ref{ppcoll1}), i.e.\ we build
conservation of both relativistic mass and linear momentum into collisions
from the outset.  Inelastic collisions are def\/ined as collisions in which
there is only one outgoing particle\footnote{As is well known, when two particles undergo a relativistic inelastic collision, kinetic energy is not conserved~\cite{Firk10}. Nonetheless, the total energy of the system~-- and hence its combined relativistic mass~-- remains unchanged \cite[\S~4.3]{dinverno}, \cite[Example~3.2]{KK10}, \cite[\S~6.2]{Rin}.}.

Informally, \ax{DYN} consists of the basic assumptions in \ax{BA} together
with the following axioms (see Fig.~\ref{fig:dynax}):

\begin{itemize}\itemsep=0pt
\item
  \ax{AxColl_3}\\
  If one observer sees a 3-particle collision, all the others see it as well.
\item
  	\ax{Ax\forall \inecoll^-}\\
  	Given any two inertial particles $a$ and $b$ with given velocities and relativistic masses, we can f\/ind a (possibly dif\/ferent) pair of particles $a'$ and $b'$ which can collide inelastically, where the relativistic masses and velocities of $a'$ and $b'$ agree with those of $a$ and $b$ respectively.
\item
	\ax{AxSpd^{-}}\\
	If two observers agree on the speed of a `slow' inertial particle, then they agree on its relativistic mass.
	
	Notice, however, that we make no assumptions concerning the relationship between relativistic mass and speed for FTL particles.  Instead, this will emerge as a logical consequence of the axioms.
\item
	\ax{AxMass}\\
	If the velocities and relativistic masses of two inertial
	particles coincide for one observer, then their relativistic masses also coincide
	for all other observers.
\item
	\ax{AxThEx^-}\\
	Given any observers $m$ and $h$, and any non-negative speed $v$ slower than that of $h$ as observed by $m$, we can assume the existence of a massive inertial particle with speed $v$ relative to~$m$.
\end{itemize}

In relativistic terms, \ax{AxThEx^-} says that inertial particles with
non-zero relativistic mass can be assumed to travel at any desired sub-light speed
(because observer $h$ cannot travel faster than light in our framework), but
the axiom is phrased so as to remain valid in Newtonian dynamics also.  The
dif\/ference between the relativistic and Newtonian systems is captured by

\begin{itemize}\itemsep=0pt
\item
  \ax{AxPh}, the \textit{light axiom}\\
  The speed of light is $1$ for every observer;
\item
  \ax{AxAbsSim}, \textit{absolute simultaneity}\\
	If two events are simultaneous for one observer, then they are simultaneous for all observers.
\end{itemize}

\noindent In our theories, relativistic dynamics is captured by
\ax{DYN}+\ax{AxPh}, while \ax{DYN}+\ax{AxAbsSim} is an axiom system for
Newtonian dynamics \cite{FTLDyn}.

Writing $\mathsf{m}_k(b)$ for the relativistic mass and $v_k(b)$ for the speed of
particle $b$ according to observer $k$, the main result of this paper
(Theorem~\ref{mainthm}) describes how particle mass (including that of
FTL particles) varies with speed in both relativistic and Newtonian dynamics.  For all inertial observers~$k$ and~$h$, and any inertial particle~$b$ moving with a f\/inite
speed according to both of them, we have

\begin{itemize}\itemsep=0pt
\item
  if relativistic dynamics (\ax{DYN}+\ax{AxPh}) holds, then
  \begin{gather}
    \mathsf{m}_k(b)\cdot \sqrt{\Abs{ 1-v_k(b)^2 }} = \mathsf{m}_h(b)\cdot \sqrt{\Abs{ 1-v_h(b)^2}};
  \label{eq:AxPh}
  \end{gather}
\item
  if Newtonian dynamics (\ax{DYN}+\ax{AxAbsSim}) holds, then
  \begin{gather*}
    \mathsf{m}_k(b) = \mathsf{m}_h(b).
  \end{gather*}
\end{itemize}

Equation \eqref{eq:AxPh} gives back the usual relativistic mass-increase theorem for slower
than light particles, but predicts that both the relativistic mass and the
momentum of an FTL particle with positive mass should decrease with speed.

It is worth noting that the axiom system \ax{DYN}+\ax{AxPh} is self-consistent because it is more general that the axiom system \ax{SRDyn} of \cite{FTLDyn}, which has been shown there to be logically consistent. Likewise, a similar construction to that used in  \cite{FTLDyn} shows that the axiom system \ax{DYN}+\ax{AxAbsSim} is also self-consistent.

\section{Formalization} \label{sec:basic-concepts}

We begin by formalising a number of basic concepts, including the axioms described above, and introduce some notation. For brevity we have adopted the familiar expositional style of formal mathematics, but it should be noted that everything can also be expressed in the pure f\/irst-order logic framework of, \eg \cite{dyn-studia,FTLDyn}.

\subsection{Quantities and bodies} We use the algebraic structure \Q
$\equiv$ \Tuple{\Q, +, \cdot, <} to represent \EMPH{quantities} (the values
used to measure relativistic masses, speed, momenta, etc.), where $+$ (addition) and
$\cdot$ (multiplication) are binary operations, and $<$ (less than) is a
binary relation, on the set $\Q$. As usual, we write $\Q^n$ for the Cartesian
product of $n$ copies of \Q, and write $x_i$ for the $i^\mathrm{th}$
component of $\vx = (x_1, \dots, x_n) \in Q^n$.

We write \B for the set of \EMPH{bodies}, and pick out three types of body
in particular: the sets \IOb of \EMPH{inertial observers}, \Ip of
\EMPH{inertial particles} and \Ph of \EMPH{photons} are subsets of~\B.
We do not assume a priori that these subsets are disjoint~-- the
fact that, e.g.,  photons cannot be inertial observers in the relativistic case, arises instead as a
theorem of the logic.

Given any inertial observers $k, h \in \IOb$ and body $b \in
\B$,
\begin{itemize}\itemsep=0pt
\item $\wl_k(b) \subseteq \Q^4$ is the \EMPH{world-line} of $b$ according
  to observer $k$;
\item $\w_{kh} : \Q^4 \rightarrow \Q^4$ is the
  \EMPH{world-view transformation} between the world-views (coordinate
  systems) of $k$ and $h$;
\item $\mathsf{m}_k(b) \in \Q$ is the
  \EMPH{relativistic mass} of $b$ according to $k$.
\end{itemize}

\subsection{Basic axioms} \label{sec:axioms}

Throughout the paper we assume a set \ax{BA} of four basic axioms in our
def\/initions and axioms without mentioning this explicitly.  Formally, we
have
\begin{gather*} 
 \ax{BA} \leteq \{ \ax{AxEField}, \ax{AxL}, \ax{AxW}, \ax{AxSelf^-}   \},
\end{gather*}
where the axioms \ax{AxEField}, \ax{AxL}, \ax{AxW} and \ax{AxSelf^-} are
def\/ined below.

\subsubsection{Notation} Our notation is essentially standard.  We write
\Rset for the f\/ield of real numbers and \leteq for def\/initional equality, so
that ``$v \leteq e$'' indicates that the value $v$ is fully determined by
evaluating an appropriate instantiation of the expression $e$.  Given any
function $f \colon X \rightarrow Y$ and $S \subseteq X$, the $f$-image of
$S$ is given by $f[S] \leteq \setopen f(x) \setmid x \in S \setclose$.

\subsubsection{Field axioms}

Einstein and his followers assumed implicitly that the structure \Q is the
f\/ield \Rset of real numbers, but in fact this imposes more logical structure
than is necessary for our deductions.  We therefore make our system more
general by assuming for~\Q only the more important algebraic properties of~\Rset.

\NEWAXIOM{\ax{AxEField}}{EField}{ The quantity part \Tuple{\Q, +, \cdot, <}
is assumed to be a Euclidean f\/ield, i.e.\ it is a linearly ordered f\/ield in the
sense of abstract algebra, equipped with the usual f\/ield operations $0$, $1$, $-$
and $/$ def\/inable from addition and multiplication, and every non-negative
element $x$ has a square root, i.e.\ $\mathop{(\forall\, x >0)}(\exists\, y)(x = y^2)$.
It is easy to see that square roots can be assumed to be non-negative when
they exist, and we def\/ine the \gyok function accordingly.}

For each $n = 1, 2, 3, \dots$, the set $Q^n$ is be assumed to carry
the normal vector space structure. Regardless of $n$, we write $\BAR{0}
\leteq (0, \dots, 0)$ for the \EMPH{origin}. The \EMPH{Euclidean length}
of a vector $\vx = (x_1, \dots, x_n)$ is the non-negative quantity
$\Abs{\vx} = \sqrt{ x_1^2 + \dots + x_n^2 }$.

For simplicity, we shall make the standard assumption throughout this paper
that $n = 4$, so that spacetime has one temporal and three spatial dimensions.
Given any $\vx = (x_1, x_2, x_3, x_4) \in \Q^4$ we call $x_1$ its
\EMPH{time component} and $(x_2,x_3,x_4)$ its \EMPH{space component}.  If
\vx has time component $t$ and space component \BAR{s}, we occasionally
abuse notation and write $\vx = (t,  \BAR{s})$.

\subsubsection{World-lines} We assume that the world-line of every inertial
particle or inertial observer is a subset of a~straight line\footnote{By a \EMPH{straight line} we mean a set $\ell
\subseteq \Q^4$ for which there exist distinct $\vx, \vy \in \Q^4$ with
$\ell = \setopen \vx + q\cdot(\vy-\vx)\: :\: q \in \Q \setclose$.} containing at
least 2 distinct points.

\NEWAXIOM{\ax{AxL}}{AxL} { For every $k \in \IOb$ and $b \in \Ip \cup \IOb$
there is a straight line $\ell$ of $\Q^4$ such that $\wl_k(b) \subseteq
\ell$, and $\wl_k(b)$ contains at least two points (elements).}

\subsubsection{World-view transformation} The world-view transformation
between any two observers is an af\/f\/ine transformation (i.e.\ a~li\-near
transformation composed with a translation) which takes the
world-line of each body according to the f\/irst observer to its world-line
according to the second observer.

\NEWAXIOM{\ax{AxW}}{AxW} {$\w_{kh}$ is an af\/f\/ine transformation and $\w_{kh}[\wl_k(b)]=\wl_h(b)$ for
every $k,h \in \IOb$ and $b\in B$.  }

\subsubsection{Self-coordinatisation}

Inertial observers consider themselves to be stationary in space (but not in
time).  This makes it easy to speak about the motion of inertial observers
since it identif\/ies observers with their time axes.

\NEWAXIOM{\ax{AxSelf^-}}{axself} {If $\vx \in \wl_k(k)$, then $(x_2, x_3,
x_4) = (0, 0, 0)$.}

\subsubsection{Auxiliary def\/initions}

The \EMPH{velocity} $\vel_k(b) \in \Q^3$
and \EMPH{speed} $v_k(b) \in \Q$ of any inertial body $b \in \Ip \cup
\IOb$ according to an inertial observer $k$ are def\/ined by
\begin{gather*}
  \vel_k(b)   \leteq   ( x_2-y_2,~x_3-y_3,~x_4-y_4) / {(x_1-y_1)} \qquad  \text{and}
	\qquad v_k(b)   \leteq   |\vel_k(b)|,
\end{gather*}
where $\vx, \vy \in \wl_k(b)$ are any points for which $x_1 \neq y_1$
(these are well def\/ined concepts whenever such points exist, since they
do not depend on the choice of $\vx$ and $\vy$).

If there are no such $\vx$ and $\vy$, then $\vel_k(b)$ and $v_k(b)$ are
undef\/ined.  Indeed, if there are distinct $\vx, \vy \in \wl_k(b)$ with $x_1
= y_1$, the worldline of $b$ is a subset of a horizontal straight line and
its speed is, from $k$'s viewpoint, `inf\/inite'.  We therefore write
$v_k(b) = \infty$ in this case, and $v_k(b)<\infty$ otherwise.  We say
that the speed of body $b$ is f\/inite (inf\/inite) according to observer $k$
provided $v_k(b)<\infty$ ($v_k(b)=\infty$), respectively.

\section{Axioms for dynamics}

Throughout this paper we write $\mathsf{m}_k(b)$ for the \EMPH{relativistic mass} of
particle $b$ according to observer $k$ -- we assume here that $\mathsf{m}$ is a
primitive construct and deduce (Theorem~\ref{mainthm}) how $\mathsf{m}_k(b)$ varies
with the relative speed of $b$ relative to $k$.

\begin{figure}[h!bt] \centering
\includegraphics{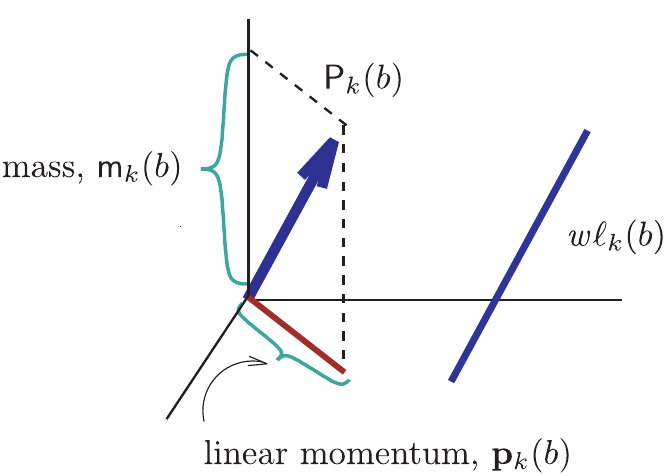}
\caption{The time component of four-momentum is relativistic mass, and its
space component is linear momentum.  Each inertial observer~$k$ considers
$b$'s four-momentum to be parallel to its world-line.} \label{fmom}
\end{figure}

We def\/ine the \EMPH{linear momentum} $\vp_k(b)$ of inertial particle $b$
according to inertial observer~$k$ in the obvious way, viz.  $\vp_k(b) =
\mathsf{m}_k(b)\cdot\vel_k(b)$.  Provided $b$ is travelling at f\/inite speed relative
to~$k$, so that $\vp_k(b)$ is def\/ined, the associated \EMPH{four-momentum}
${\fvp_k(b)}$ is the vector in $\Q^4$ whose time component is the
relativistic mass, and whose space component is the linear momentum~-- it is
not dif\/f\/icult to prove that $\fvp_k(b)$ is parallel to the world-line of $b$
(Fig.~\ref{fmom})
\begin{gather*} 
\fvp_k(b) = \begin{cases}
     \left( \mathsf{m}_k(b), \vp_k(b) \right) & \text{if $\sqspeed_k(b) < \infty$},
     \\ \mathrm{undef\/ined} & \text{otherwise}. \end{cases}
\end{gather*}

\subsection{Particles involved in a collision}

We say that an inertial particle $b$ is \EMPH{incoming} at some coordinate
point $\vx$, according to inertial observer $k$, provided $\vx \in \wl_k(b)$
and $x_1$ is an upper bound for the time coordinates of points on~$\wl_k(b)$ and the speed of $b$ is f\/inite,
i.e.\ if $\vy \in \wl_k(b)$ and $\vy \neq \vx$ then $y_1 < x_1$.
We write $\inc_k(b)$ to mean that there is some point $\vx$ at which $b$ is
incoming according to $k$, and $\inc_k(b)@\vx$ if we wish to highlight some
such \vx explicitly.  \EMPH{Outgoing} particles, $\out_k(b)$ and
$\out_k(b)@\vx$ are def\/ined analogously.

A collection $b_1,\ldots, b_n$ of inertial particles form a \EMPH{collision} according to observer $k$ provided
\begin{itemize} \itemsep=0pt
\item
	there is a point \vx at which they are all either incoming or
outgoing with f\/inite speed according to~$k$; and \item
	relativistic mass and linear momentum are both conserved at \vx
	according to $k$ (Fig.~\ref{ppcoll1}), i.e.\ allowing $b_i$ to range
	over the particles, we have \[
	  \sum_{ \{ b_i : \inc_k(b_i)@\vx\} }{\fvp_k(b_i)} = \sum_{ \{ b_i :
	    \out_k(b_i)@\vx\} }{\fvp_k(b_i)}. \]
\end{itemize}
We are particularly interested below in \EMPH{collisions involving three particles},
and write $\coll_k(abc)$ to mean that $a$, $b$, $c$
form a collision according to~$k$.

\subsubsection{Collision axioms} For each natural number $n$ we
introduce an axiom saying that the existence of any $n$-body collision is
observer-independent.  Thus, conservations of relativistic mass, linear
momentum and four-momentum do not depend on the inertial observer.

\begin{figure}[hbtp] \centering
\includegraphics{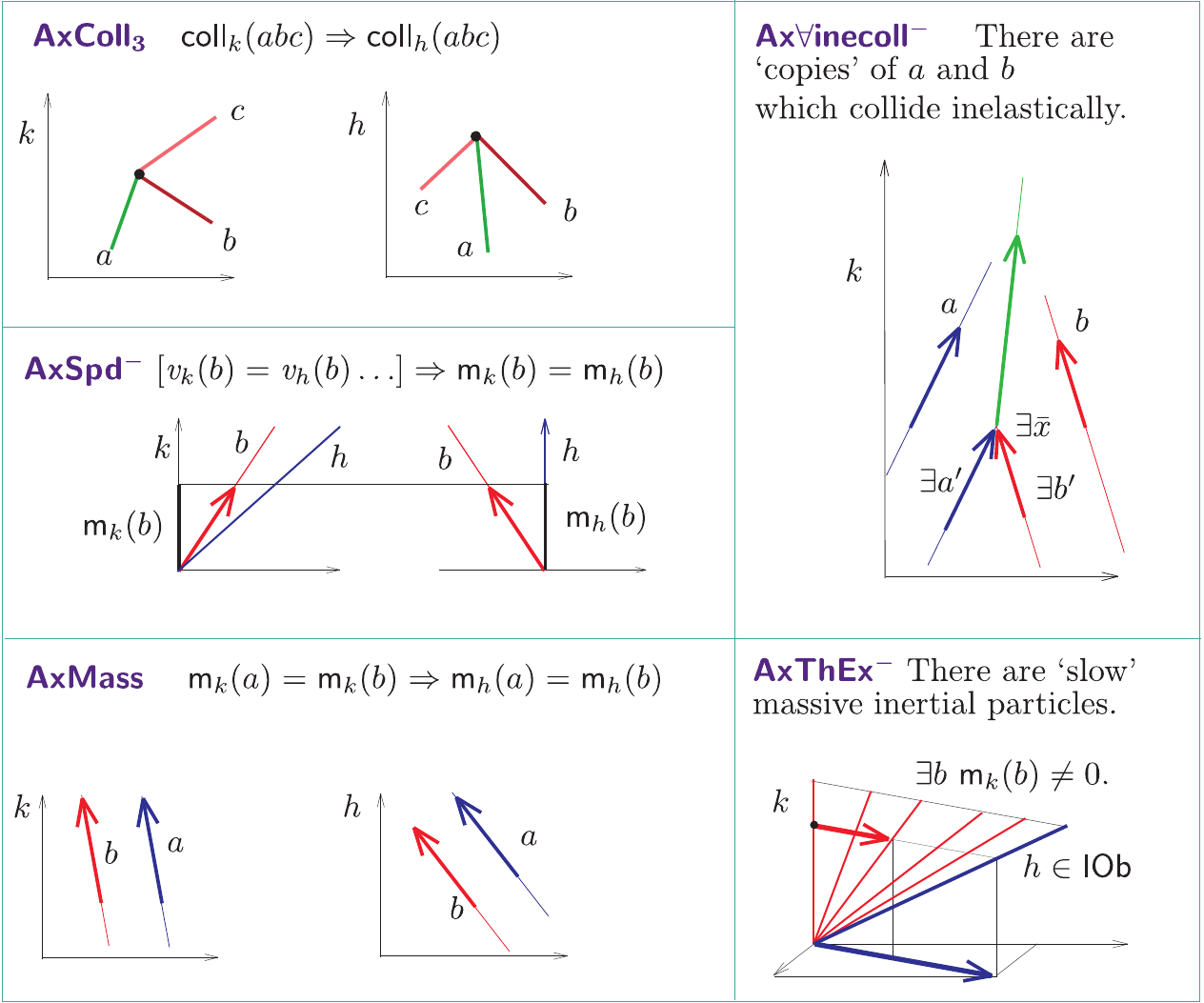}
\caption{The dynamical axioms in \ax{DYN}.} \label{fig:dynax}
\end{figure}

\NEWAXIOM{\ax{AxColl_n}}{coll} {If inertial particles $b_1, \dots, b_n$
form a~collision for one inertial observer, they also form a~collision for every other inertial observer who considers them all
to have f\/inite speed (see Fig.~\ref{fig:dynax}).}

\subsubsection{Inelastic collisions}

We say that inertial particles $a$ and $b$  collide \EMPH{inelastically}
according to observer $k$ provided there exist some inertial
particle $c$ such that $a$, $b$, $c$ form a collision,
where $a$ and $b$ are incoming and $c$ is outgoing, i.e.
\begin{gather*}
   \exists\, c\, [ \coll_k(abc) \land \inc_k(a)
\land \inc_k(b) \land \out_k(c) ].
\end{gather*}

The following axiom is slightly subtle. In general, an observer would not
expect two arbitrarily selected particles $a$ and $b$ to be following
worldlines that necessarily meet.  We can, however, f\/ind paths
\EMPH{parallel to} those followed by $a$ and $b$ (i.e.\ followed by particles
with the same four-momenta), which meet at some point.  The axiom states
this formally, and says that we can form an inelastic collision of these
parallel-moving particles at \vx, provided they have `sensible' relativistic masses and
speeds according to the observer in question. In other words, this axiom says that all `sensible' collisions are realisable.

\NEWAXIOM{\ax{Ax\forall\inecoll^-}}{axinecoll} {If $k \in \IOb$ and $a, b
\in \Ip$ satisfy $v_k(a) < \infty$, $v_k(b) < \infty$ and $\mathsf{m}_k(a) +
\mathsf{m}_k(b) \neq 0$, then there exist $a', b' \in \Ip$ with $\fvp_k(a') =
\fvp_k(a)$ and $\fvp_k(b') = \fvp_k(b)$, such that $a'$ and $b'$ collide
inelastically.  }

We typically use these collisions to determine relationships between the
four-momenta of $a$ and $b$, but of course it is actually $a'$ and $b'$ that
take part in the collision.  Nonetheless our deductions remain valid,
because $a$ and $a'$ have the same four-momentum, as do $b$ and $b'$, and these equalities continue to hold in every other observer's worldview by \ax{AxMass} below.

\subsubsection{Relativistic mass}

Given $\bu, \bv \in \Q^{3}$, we write $\bu
\preceq \bv$ to mean that $\bu$ and $\bv$ point in the same direction, and
the length of $\bu$ is less than or equal to the length of~$\bv$, i.e. \[
  \bu \preceq \bv \quad \defiff \quad \exists\, \lambda\;  (0 \leq
  \lambda \leq 1 \lland \bu = \lambda \cdot \bv ).
\]

The following axioms concern the relativistic masses of slow particles, and are
expressed in a~way that makes them equally meaningful for both relativistic
and Newtonian dynamics.  Since inertial observers travel slower than light in the relativistic setting, but can achieve
any sublight speed we wish, we can def\/ine a particle to be `slow' provided
there is some observer who appears to be travelling faster than it.  That
is, observer~$k$ considers body~$b$ to be `slow' provided there is some
other observer~$h$ for which $\vel_k(b) \preceq \vel_k(h)$.

Our next axiom asserts that whenever two observers agree on the speed of a slow inertial particle, then they also agree on its relativistic mass.  That is, suppose observer~$k$ considers inertial particle~$b$ and
observer~$h$ to be moving in the same direction, with $h$ moving faster than~$b$.  If~$h$ and~$k$ see~$b$ to be moving at the same speed, they will also
agree on its relativistic mass (Fig.~\ref{fig:dynax}).

\NEWAXIOM{\ax{AxSpd^-}}{axspeed} { If $\vel_k(b) \preceq \vel_k(h)$ and
$v_k(b) = v_h(b)$, then $\mathsf{m}_k(b)=\mathsf{m}_h(b)$,  for all $k, h \in \IOb$ and $b \in \Ip$. }

It is important in axiom \ax{AxSpd^-} that particle $b$ is slow, because
in special relativity,
if observers $k$ and $h$ move with respect to one another and $b$ is
moving with the speed of light in the same spatial direction as $h$
according to $k$, then the speed
of $b$ is the same for $k$ and $h$ but its relativistic mass is dif\/ferent.
In both relativistic and Newtonian axiom systems, we can prove that no
observer can travel with inf\/inite speed, and moreover in the relativistic
axiom system every observer travels with slower than light speed, see
Corollary~\ref{corollary}. 

We also assume that if the relativistic masses and velocities of two
inertial particles coincide for one inertial observer, then their relativistic masses
also coincide for all other inertial observers.

\NEWAXIOM{\ax{AxMass}}{axmass} { If $\fvp_k(a) = \fvp_k(b)$, then $\mathsf{m}_h(a) =
\mathsf{m}_h(b)$ for all $k, h \in \IOb$ and $a, b \in \Ip$.  }

Finally, we assume the existence of `slow' inertial particles with nonzero
relativistic masses.  That is, given any inertial observer and any `slow'
velocity $\bv$, there is an inertial particle with nonzero relativistic mass
that moves with velocity $\bv$.

\NEWAXIOM{\ax{AxThEx^-}}{axthexp} { For all $k, h \in \IOb$ and $\bv \preceq
\vel_k(h)$ there is $b \in \Ip$ satisfying $\vel_k(b) = \bv$ and $\mathsf{m}_k(b)
\neq 0$.  }

\subsubsection{Shared axiom system for dynamics} We can now def\/ine the axiom
system \ax{DYN} for dynamics, which is common to both the relativistic and
Newtonian settings.  It comprises the basic assumptions, together with the
axioms introduced above, viz.  \[
  \ax{DYN}  \leteq \ax{BA} \cup \{\ax{AxColl_3},
               \ax{Ax\forall \inecoll^-}, \ax{AxSpd^-}, \ax{AxMass},
               \ax{AxThEx^-}\}.
\]

\subsubsection{Dif\/ferentiating between relativistic and Newtonian dynamics}

\begin{figure}[t] \centering
\includegraphics{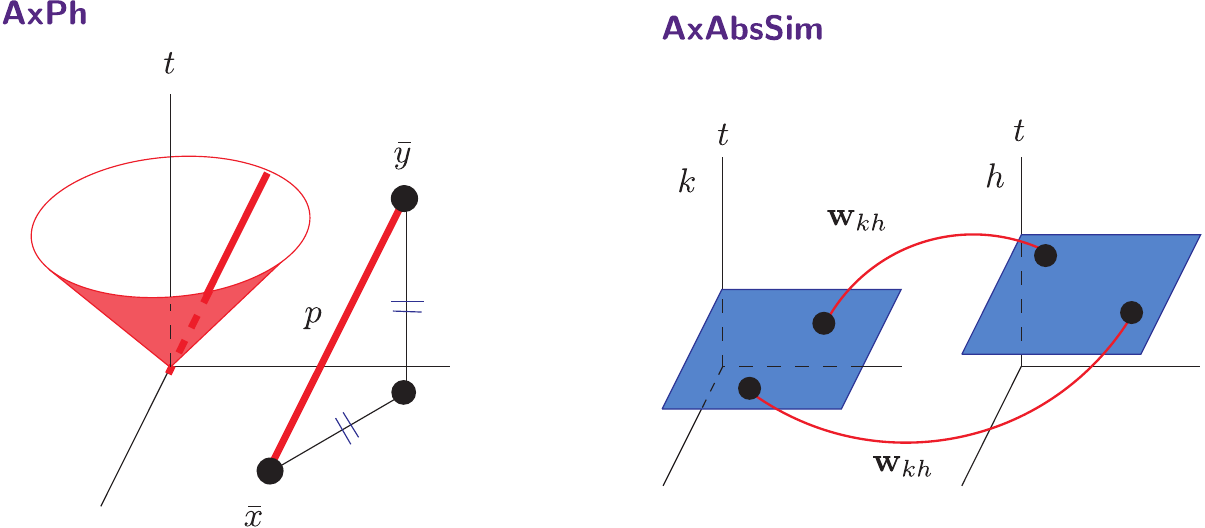}
\caption{Illustration for the photon axiom and for the absolute simultaneity
axiom.} \label{phabss}
\end{figure}

Finally, we def\/ine the axioms that allow us to distinguish between the two
systems (Fig.~\ref{phabss}).

The \EMPH{photon axiom}, \ax{AxPh}, says that each inertial observer
considers the speed of light to be $1$ everywhere and in every direction (in
particular, therefore, it is f\/inite, whence this axiom characterises
relativistic dynamics).  Moreover, it is always possible to send out such
light signals, i.e.\ two points lie on a photon's worldline if and only if the
slope of the line joining them is~$1$.

\NEWAXIOM{\ax{AxPh}}{axph}
{ For every inertial observer $k \in \IOb$, \[
  (\exists\, p \in \Ph)(\vx,\vy\in\wl_k(p)) \quad \Leftrightarrow \quad
  (x_2-y_2)^2 + (x_3-y_3)^2 +(x_4-y_4)^2 = (x_1-y_1)^2 .
\]
}

In contrast, the \EMPH{absolute simultaneity} axiom, \ax{AxAbsSim}, which
characterises Newtonian dynamics, says that whenever two events are
simultaneous for one observer then they are simultaneous for every observer.

\NEWAXIOM{\ax{AxAbsSim}}{AxAbsSimym}
{
If $\w_{kh}(\vx)=\vx'$ and $\w_{kh}(\vy)=\vy'$, then $x_1=y_1 \Leftrightarrow x'_1=y'_1$ for every $k, h \in \IOb$.
}

The axiom system $\ax{DYN}\cup\{\ax{AxPh}\}$ is our axiom system for
relativistic dynamics, and axiom system $\ax{DYN}\cup\{\ax{AxAbsSim}\}$ is our
axiom system for Newtonian dynamics.

\section{The main theorem and its proof} \label{sec:main-theorem}

We can now state and prove our main theorem.
Our central claim, concerning the way that relativistic masses transform as speed
increases, then follows as an easy corollary.

\begin{theorem} \label{mainthm} Assume axiom system \ax{DYN}.  Let $b$ be an
inertial particle and let $k$, $h$ be inertial observers for which the speed
of $b$ is finite.  Then{\samepage
\begin{itemize}\itemsep=0pt
\item[$(a)$] $\mathsf{m}_k(b)\cdot \sqrt{\Abs{1-v_k(b)^2}} = \mathsf{m}_h(b)\cdot \sqrt{\Abs{1-v_h(b)^2}}$ if \ax{AxPh}
holds;
\item[$(b)$] $\mathsf{m}_k(b)=\mathsf{m}_h(b)$ if \ax{AxAbsSim} holds.
\end{itemize}}
\end{theorem}

To see how positive FTL mass and momentum change in the
relativistic setting depending on the particle's speed relative to an
inertial observer, let us assume that $b$ is an FTL particle relative to $k$, and that $v_h(b) < v_k(b)$.  It is known
\cite{SN12} that the axioms assumed here prohibit inertial observers (though not FTL inertial particles in general) from travelling FTL with respect to one another, so we can
also assume that $b$ is travelling FTL relative to $h$.  Since both
observers consider $v(b)$ to be greater than $1$, the theorem tells us
that \[
    \mathsf{m}_k(b) = \mathsf{m}_h(b) \cdot \sqrt{ \frac{ v_h(b)^2 - 1 } { v_k(b)^2 - 1
     } } <
                           \mathsf{m}_h(b),
                            \]
so that relativistic mass is considered to be lower by the observer who is moving
faster relative to the particle.  Similarly, the observers' momentum
measurements satisfy \[
    \Abs{\vp_k(b)} = \mathsf{m}_k(b) \cdot v_k(b) = \mathsf{m}_h(b) \cdot \sqrt{ \frac{
     v_h(b)^2 - 1 }
                           { v_k(b)^2 - 1 } } \cdot v_k(b)   .
\] A simple rearrangement of terms shows that \[
  v_k(b) > v_h(b) \quad \Rightarrow \quad \sqrt{ \frac{ v_k(b)^2\cdot
     \left( v_h(b)^2 - 1 \right) } { v_k(b)^2 -
                           1 } } < v_h(b), \] and hence \[
    \Abs{\vp_k(b)} = \mathsf{m}_k(b) \cdot v_k(b) < \mathsf{m}_h(b) \cdot v_h(b) =
     \Abs{\vp_h(b)}, \]
so that  momentum also appears to drop as the particle's speed
increases.

\begin{figure}[t]
\centering
\includegraphics{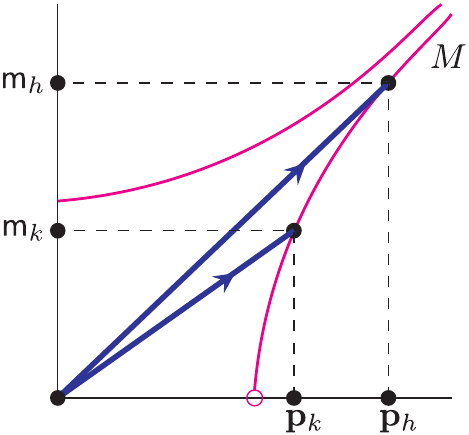}
\caption{A spacetime diagram showing the relativistic mass $\mathsf{m}$ and linear momentum~$\vp$ of a positive-mass FTL body $b$ as seen by two observers $k$ and $h$ in the relativistic setting. The arrows represent the four-momenta of $b$ as seen by each observer, and the continuous lines show the Minkowski circle on which their tips are constrained to lie.}
\label{fig-decrease} \end{figure}

We can also interpret this result as shown in Fig.~\ref{fig-decrease}. Notice
that $\mathsf{m}_k(b)\cdot \sqrt{\Abs{1-v_k(b)^2}}$ is essentially the Minkowski norm \NormM{\fvp_k(b)} of $b$'s four-momentum as seen by $k$ (see Section~\ref{def:minkowski-norm} for the def\/inition of \NormM{\cdot}), except that the actual, rather than absolute, value of $\mathsf{m}_k(b)$ is used. Thus the theorem states that the two observers agree both on the Minkowski length of $\fvp(b)$, and on the sign they assign to $b$'s relativistic mass.  As illustrated in the diagram, the points $\fvp_k = (\mathsf{m}_k, \vp_k)$ and $\fvp_h = (\mathsf{m}_h, \vp_h)$ must, therefore, lie on the same component of the Minkowski sphere $M$, and since~$b$ is faster for $k$ than for $h$ the corresponding four-momentum is closer to the horizontal, whence it is clear that $\mathsf{m}_k < \mathsf{m}_h$ and $\Abs{\vp_k} < \Abs{\vp_h}$.

\subsection*{Outline of the proof}

Let \Norm{\cdot} be either of the norms \NormM{\cdot} or \NormT{\cdot},
where the \EMPH{Minkowski norm} \NormM{\vx} and \EMPH{time norm}~\NormT{\vx}
of $\vx \in \Q^4$ are given by \[
  \NormM{\vx} = \sqrt{\Abs{x_1^2-x_2^2-x_3^2 -x_4^2}} \qquad \text{and}
	\qquad \NormT{\vx} = \Abs{x_1}.  \]
\label{def:minkowski-norm}

Recall that a linear transformation $L : \Q^4 \rightarrow \Q^4$ is called a
\EMPH{Lorentz transformation} if it preserves the Minkowski norm.  We will
say that a linear transformation is \EMPH{time-preserving} if it preserves
the time-norm.  An \EMPH{expansion} on $\Q^4$ is a mapping $E_q : \Q^4
\rightarrow \Q^4$ given by $E_q(\vx) \leteq q \cdot \vx$, where $0 < q \in
\Q$.

The proof requires four basic lemmas.
\begin{enumerate}\itemsep=0pt
\item
    Given two observers $k$ and $h$ moving with f\/inite (possibly zero) speed with respect to one another, we show (Lemma~\ref{lem:special-particle}) that there exists a slowly moving massive inertial particle $a$ on whose speed they agree (doing so f\/irst requires us to show (Lemma~\ref{lem:massless}) that, subject to certain speed restrictions, we have $\mathsf{m}_h(b) = 0$ whenever $\mathsf{m}_k(b) = 0$).  According to~\ax{AxSpd^-}, the two observers therefore agree on the particle's relativistic mass as well, and hence \Norm{\fvp_k(a)} = \Norm{\fvp_h(a)}.
\item
    To prove the main theorem, we consider a collision between $a$ and an arbitrary massive particle $b$, and use basic linear algebra to deduce both that \Norm{\fvp_k(b)} = \Norm{\fvp_h(b)} and that~$k$ and~$h$ agree as to the sign of $b$'s relativistic mass, where the choice of norm depends on whether we are assuming \ax{AxPh} or \ax{AxAbsSim} in addition to \ax{DYN}.  In order to do this, we f\/irst need to prove that the linear part of the worldview transformation can be written as a~composition of a norm-preserving transformation $L$ and an expansion~$E_q$  (Lemma~\ref{lem:decomposition}), and show under what conditions two observers assign the same sign to a particle's relativistic mass (Lemma~\ref{lem:aux}).
\end{enumerate}

\begin{lemma} \label{lem:massless} Assume $\ax{DYN} \setminus
\{\ax{AxSpd^-}\}$, and suppose $k, h \in \IOb$ and $b \in \Ip$ satisfy
$v_k(b) < \infty$, $v_h(b) < \infty$, $v_k(h) < \infty$, and
$v_h(k)<\infty$.  Then $\mathsf{m}_k(b) = 0$ if and only if $\mathsf{m}_h(b) = 0$.
\end{lemma}

\begin{proof} Suppose $\mathsf{m}_k(b) = 0$.  We will prove that $\mathsf{m}_h(b) = 0$ as
well.  The converse will follow by symmetry.

Axiom \ax{AxThEx^-} says we can choose an inertial particle $a$ such that
$\mathsf{m}_k(a) \neq 0$, $v_k(a) < \infty$ and $v_h(a) < \infty$.  By
\ax{Ax\forall\inecoll^-}, we can collide $a$ and $b$ inelastically as
incoming particles (from $k$'s point of view) to form an outgoing particle
$c$, and we have $\fvp_k(a)+\fvp_k(b)=\fvp_k(c)$.  Because $\mathsf{m}_k(b)=0$, we
know that $\fvp_k(b)\leteq(\mathsf{m}_k(b),\mathsf{m}_k(b)\cdot\vel_k(b))=\bar 0$, whence
$\fvp_k(a)=\fvp_k(c)$ and $k$ considers $a$ and $c$ to have parallel
worldlines.  Axiom \ax{AxMass} now tells us that $\mathsf{m}_h(a) = \mathsf{m}_h(c)$, and
since $\w_{kh}$ is af\/f\/ine, it maps parallel worldlines to parallel
worldlines, so $h$ considers $a$ and $c$ to have identical velocities.  By
def\/inition, it follows that $\fvp_h(a)=\fvp_h(c)$.

We know that $a$, $b$ and $c$ form a collision for $k$, so they also form
one for $h$ by \ax{AxColl_3}.  We know that $a$ and $c$ have parallel
worldlines from $k$'s point of view, so, since they coincide at the
collision point, the same straight line contains both of them in $k$'s
worldview, with the collision point marking a clear boundary between one
worldline and the other.  Since $\w_{kh}$ is af\/f\/ine, it takes straight lines
to straight lines without changing the delimiting nature of the `boundary
point', so according to $h$, one of $a$ and $c$ is incoming and the other
one is outgoing.  Since they form a collision, we conclude that either \[
    \fvp_h(a) + \fvp_h(b) = \fvp_h(c) \qquad \text{or} \qquad \fvp_h(a) =
\fvp_h(c) + \fvp_h(b) \] depending on whether $b$ is seen to be incoming or
outgoing by $h$.  In either case it follows that $\fvp_h(b)= \bar 0$, whence
$\mathsf{m}_h(b) \equiv (\fvp_h(b))_1 = 0$ as claimed.
\end{proof}

The next lemma shows that two observers, moving at f\/inite speed relative to
one another, can identify a particle $b$ on whose speed they agree.  The
proof outline given below makes a~simplifying assumption -- equation~\eqref{felteves} -- and assumes that a version of Bolzano's theorem holds for~\Q (i.e.\ if a continuous \Q-valued function on~$[0,1]$ takes both positive and
negative values, then it must also take the value 0 at some point in the
interval).  A fully rigorous proof of the lemma is provided in the Appendix.

\begin{lemma} \label{lem:special-particle} Assume \ax{BA} and \ax{AxThEx^-},
and suppose $k$ and $h$ are inertial observers satisfying $v_k(h) <
\infty$ and $v_h(k) < \infty$.  Then there exists an inertial particle $b$
such that $\vel_k(b) \preceq \vel_k(h)$, $v_k(b)=v_h(b)$ and
$\mathsf{m}_k(b)\neq 0$.  \end{lemma}

\begin{proof}[Proof outline] If $v_k(h) = 0$, then by \ax{AxThEx^-}, there
is $b \in \Ip$ such that $v_k(b) = 0$ and $\mathsf{m}_k(b) \neq 0$.  This $b$ has
the desired properties, so without loss of generality we shall assume that
$v_k(h) > 0$.

For simplicity assume that, for every inertial particle $b$,
\begin{gather} \label{felteves}
    v_k(b) \leq v_k(h) \quad \Rightarrow \quad v_h(b) < \infty.
\end{gather} This implication holds automatically if we assume either
\ax{AxPh} or \ax{AxAbsSim} (see Corollary~\ref{corollary}), as we do in the
proof of the main theorem, but is not strictly necessary in the current
context; see the Appendix.

For each $x\in[0,1]$, choose $b_x \in \Ip$ such that $\vel_k(b_x) = x \cdot
\vel_k(h)$ and $\mathsf{m}_k(b_x) \neq 0$, and note that $b_x$'s existence is
guaranteed by \ax{AxThEx^-}.  Since \eqref{felteves} implies that
$v_h(b_x) < \infty$, we can def\/ine \[
	h(x) = v_k(b_x) - v_h(b_x) \] and it is easy to see that $h$ is
continuous on [0,1].  By $v_k(b_0)=0$, $v_h(b_0)>0$, $v_k(b_1)>0$ and
$v_k(b_1)=0$, we have that $h(0) < 0$ and $h(1)>0$.  Now Bolzano's theorem
tells us there is some $u \in [0,1]$ such that $h(u) = 0$, i.e.\ $v_k(b_u) =
v_h(b_u)$.  Taking $b = b_u$ completes the proof.  \end{proof}

\begin{lemma} \label{lem:decomposition} Assume \ax{BA} and suppose $k, h \in
\IOb$.  There exists a linear transformation $L$ and an expansion $E_q$ such
that the linear part of $\w_{kh}$ is the composition $E_q \circ L$ of $L$
with $E_q$, viz.{\samepage
  \begin{gather*} 
  \w_{kh}(\vx) - \w_{kh}(\bar 0) = q \cdot
    L(\vx) \quad \text{for all $\vx \in \Q^4$}, \quad \text{where}
 \end{gather*}
\begin{itemize}\itemsep=0pt
  \item[$(a)$] if \ax{AxPh} holds, $L$ is a Lorentz transformation; \item[$(b)$]
  if \ax{AxAbsSim} holds, $L$ is a time-preserving transformation.
\end{itemize}}
\end{lemma}

\begin{proof}
   (a) Notice f\/irst that $\w_{kh}$ is necessarily `photon preserving', in
that it maps any line of slope one in $k$'s worldview to a line of slope one
in $h$'s worldview.  To see why, suppose $\ell_k$ is such a line.  By
\ax{AxPh} we can choose a photon $p$ whose worldline contains two points
within $\ell_k$.  Writing $\ell_h \leteq \w_{kh}(\ell_k)$, we know that
$\ell_h$ is again a straight line (since $\w_{kh}$ is assumed af\/f\/ine, by~\ax{AxW}).  Since $p$ is a photon and $\ell_h$ contains two points at which
$p$ is present from $h$'s point of view, it follows that the slope of
$\ell_h$ is also equal to 1 (again by \ax{AxPh}).  But in
\cite[Theorem~3.6.4]{pezsgo} it is shown that any photon-preserving af\/f\/ine
transformation can be written as the composition of a Poincar\'e
transformation and an expansion, and the result follows.

	(b) Notice f\/irst that the linear part (call it $\Phi$) of $\w_{kh}$
maps the hyperplane $x_1 = 0$ to itself, since $\Phi(0,\BAR{s})$ has to be
simultaneous with $\Phi(\BAR{0}) = \BAR{0}$ for all $\BAR{s} \in \Q^3$ by~\ax{AxAbsSim}.  Since $\Phi$ is linear, it therefore maps the hyperplane
$x_1 = 1$ to some hyperplane $x_1 = \lambda$, where $\lambda \in \Q$ is
non-zero.  It follows that $\Phi$'s action can be written in the form
$\Phi(x_1, \BAR{s}) = (\lambda \cdot x_1,\; \phi(\BAR{s}))$, where $\phi$ is
a linear map on $\Q^3$.  Def\/ining $q = \Abs{\lambda}$ ensures that $q > 0$,
so that the expansions $E_q$ and $E_{1/q}$ are def\/ined.  If we now def\/ine $L
= E_{1/q} \circ \Phi$, it follows immediately that $L$ is time-preserving,
and since $\Phi = E_q \circ (E_{1/q} \circ \Phi) = E_q \circ L$, we are
done.  \end{proof}

An easy corollary (since every inertial observer $h$ considers that $v_h(h) = 0$ by \ax{AxSelf^-}) is that there are no FTL inertial observers in the relativistic case and no inertial observers moving with inf\/inite speed in the Newtonian case.

\begin{corollary} \label{corollary} Assume \ax{BA} and suppose $k,h\in\IOb$.
Then $v_k(h)<1$ if \ax{AxPh} holds, and $v_k(h)<\infty$ if \ax{AxAbsSim}
holds.
\qed
\end{corollary}

Given any inertial observer $k$ and incoming or outgoing inertial particle
$b$, write \[
  \fva_k(b) = \begin{cases} \fvp_k(b) & \text{if $b$ is incoming}, \\
	  -\fvp_k(b) & \text{if $b$ is outgoing}. \end{cases} \]

\begin{lemma} 
\label{lem:aux} Assume \ax{BA}.  Let $k$ and $h$
be inertial observers moving at finite speed relative to one another, and
let $b$ be an inertial particle which $k$ considers to be either incoming or
outgoing at some point and that $v_h(b)<\infty$.  Suppose also that the
linear part of $\w_{kh}$ is a composition $E_q \circ L$, where $E_q$ is an
expansion $(q>0)$ and $L$ is a linear transformation.

Then \begin{itemize}\itemsep=0pt
\item[$(a)$] if $\fva_h(b) = L(\fva_k(b))$, then $\mathsf{m}_k(b)
\cdot \mathsf{m}_h(b) \geq 0$;
\item[$(b)$] if $\fva_h(b) = -L(\fva_k(b))$, then
$\mathsf{m}_k(b) \cdot \mathsf{m}_h(b) \leq 0$.  \end{itemize}
\end{lemma}

\begin{proof} Assume without loss of generality that $b$ is incoming for
$k$.  Then $\fva_k(b)=\fvp_k(b)$.  The proof when $b$ is outgoing is
analogous.  Moreover, if $\mathsf{m}_k(b) = 0$, the claims follow  trivially, so
assume without loss of generality that $\mathsf{m}_k(b)>0$.

(a) If $b$ is incoming for $h$, then $\fva_h(b) = \fvp_h(b)$ and so
$\fvp_h(b) = L(\fvp_k(b))$.  Since $b$ is incoming for $h$ we know that its
worldline exists to the past of the collision point, and likewise for $k$,
so $L$ cannot have reversed the temporal sense of $b$'s worldline, and hence
cannot have changed the sign of the f\/irst component of $\fvp_k(b)$.  On the
other hand, if $b$ is outgoing for $h$, then $\fva_h(b) = - \fvp_h(b)$ and
so $\fvp_h(b) = - L(\fvp_k(b))$.  This time we know that $L$ must have
reversed the temporal sense of $b$'s worldline, since it maps a worldline to
the past of the collision event to one that exists in its future.  This is
then reversed again by the negative sign in ``$- L(\fvp_k(b))$''.  In either
case, we therefore f\/ind that the time-components of $\fvp_k(b)$ and
$\fvp_h(b)$ must have the same sign (see Fig.~\ref{lemma4fig}).  Since this
component is the relativistic mass, we must have $\mathsf{m}_h(b) > 0$, and the
result follows.
\begin{figure}[hbtp] \centering
\includegraphics{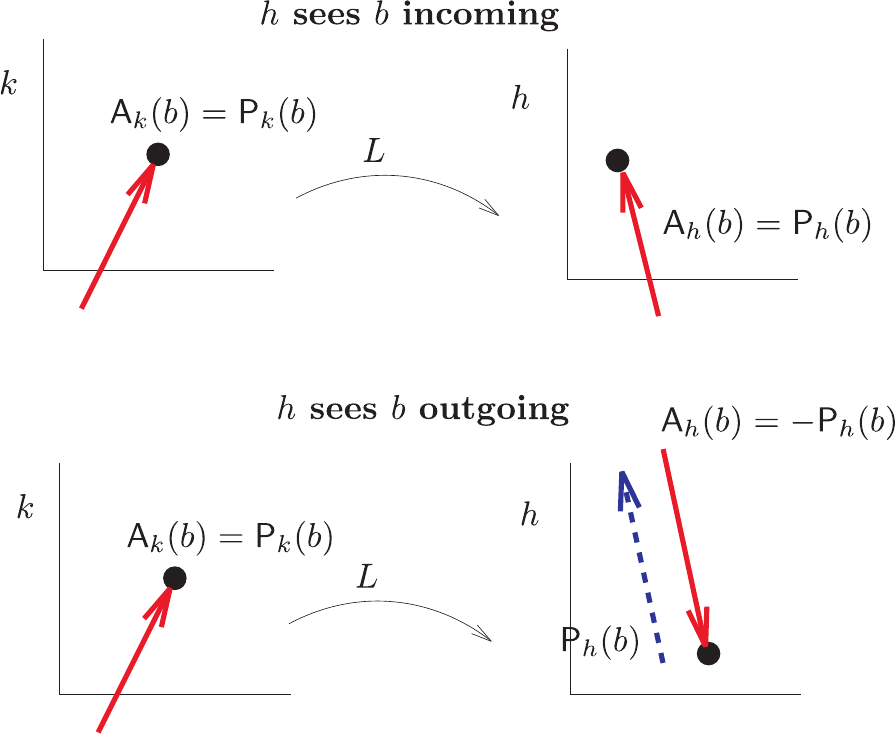}
\caption{Illustration for the proof of~(a) of Lemma~\ref{lem:aux}.} \label{lemma4fig}
\end{figure}

(b) The proof of this part is similar. If $b$ is incoming for $h$, then $L$
does not reverse the temporal sense of $b$'s worldline, but nonetheless
$\fvp_h(b) = -L(\fvp_k(b))$.  On the other hand, if $b$ is outgoing for $h$,
then $\fvp_h(b) = L(\fvp_k(b))$ and $L$ has `reversed time' (it has switched
an incoming particle to an outgoing one).  In either case, we f\/ind that the
time-components of~$\fvp_k(b)$ and~$\fvp_h(b)$ must have opposite signs, and
the result again follows.  \end{proof}

\begin{proof}[Proof of main theorem] Suppose $k$ and $h$ are inertial
observers and recall that, by Corollary~\ref{corollary}, all inertial observers move
at f\/inite speed relative to one another, and that by Lemma
\ref{lem:special-particle}, there exists a massive slow inertial
particle $a$ on whose speed they agree.

Let $b$ be an arbitrary inertial particle. We will prove the theorem holds for $b$.  Form a collision between $a$ and
$b$, and call the resulting outgoing particle $c$.  Without loss of
generality we can assume that $c$ has f\/inite speed relative to both $k$ and
$h$, for if not we can use \ax{Ax\forall \inecoll^-} to replace $a$ with
multiple copies of itself to change the resulting trajectory of $c$.  Then
by \ax{AxColl_3} we have that $a$, $b$, $c$ form a collision according to
$h$.  There are three cases to consider.

(a) If $\mathsf{m}_k(b) = 0$, then we know from Lemma \ref{lem:massless} that
$\mathsf{m}_h(b) = 0$, and so the theorem holds.

(b) Suppose $\mathsf{m}_k(b) \neq 0$. If $\vel_k(b) = \vel_k(a)$, then because $\w_{kh}$ is af\/f\/ine, we also
have $\vel_h(b) = \vel_h(a)$ and hence, by choice of $a$, that $v_k(b) =
v_h(b)$.  By \ax{AxSpd^-} it now follows that $\mathsf{m}_k(b) = \mathsf{m}_h(b)$.  So $k$
and $h$ agree on both $b$'s speed and its relativistic mass, and the theorem
again holds.

(c) Suppose $\mathsf{m}_k(b) \neq 0$ and $\vel_k(b) \neq
\vel_k(a)$ (Fig.~\ref{prooffig}).  Then the worldlines of $a$ and $b$ are
not parallel in $k$'s worldview, so their four-momenta cannot be linearly
dependent.  We also know that $\fvp_k(a) + \fvp_k(b) = \fvp_k(c)$ because
$a$, $b$ and $c$ form a collision.  It follows that no two of the vectors
$L(\fvp_k(a))$, $L(\fvp_k(b))$ and $L(\fvp_k(c))$ can be linearly dependent.
\begin{figure}[hbtp] \centering
\includegraphics{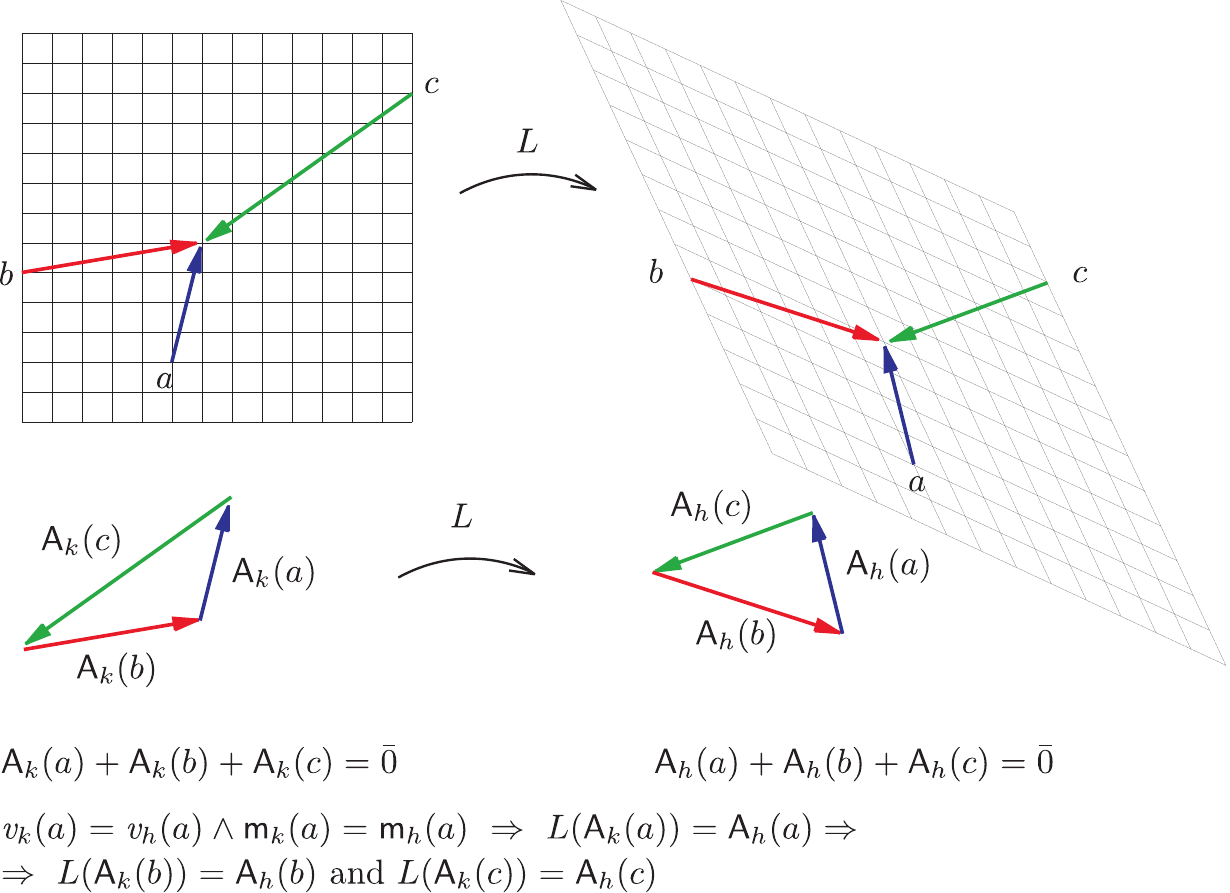}
\caption{Illustration for the proof of the main theorem.} \label{prooffig}
\end{figure}

Using the notation of Lemma~\ref{lem:aux}, the collision conditions on $a$,
$b$ and $c$ can be written
\[ \fva_k(a) + \fva_k(b) + \fva_k(c) = \BAR{0} = \fva_h(a) + \fva_h(b) +
\fva_h(c) \] and hence \begin{gather}
   (\fva_h(a) - L(\fva_k(a))) + (\fva_h(b) - L(\fva_k(b))) + (\fva_h(c) -
L(\fva_k(c))) = \BAR{0} .  \label{eqn:proof} \end{gather}

For each particle $d$, however, we know that $L(\fva_k(d))$ is parallel to
$\fva_h(d)$, because each of these is parallel to $d$'s worldline as seen by
$h$.  We claim that $\fva_h(a) = L(\fva_k(a))$.  We have already observed
that $L(\fva_k(a))$ is parallel to $\fva_h(a)$.  To see that they have the
same size (whether in the Minkowski norm, time norm or Euclidean norm),
recall that (since $a$ is slow), $k$~and~$h$ agree on $a$'s speed, and also
therefore (by \ax{AxSpd^-}) on its relativistic mass.  It now follows that~$k$ and~$h$ agree on the size of $\fva(a)$, and since $L$ is norm-preserving
$\fva_h(a)$ and $L(\fva_k(a))$ have the same size.
It only remains to show that $\fva_h(a)$ and $L(\fva_k(a))$
point in the same direction.  But this is immediate from Lemma
\ref{lem:aux}(b), since we would otherwise require $\mathsf{m}_k(a) \cdot \mathsf{m}_h(a)
\leq 0$, which is impossible (we have already established that they are
equal and non-zero).

It now follows from \eqref{eqn:proof} that $(\fva_h(b) - L(\fva_k(b))) +
(\fva_h(c) - L(\fva_k(c))) = \BAR{0}$, and hence by linear independence that
\[
    \fva_h(b) = L(\fva_k(b)) \qquad \text{and} \qquad \fva_h(c) =
L(\fva_k(c)) .  \] Lemma~\ref{lem:aux}(a) now tells us that $\mathsf{m}_k(b) \cdot
\mathsf{m}_h(b) \geq 0$, while Lemma \ref{lem:massless} shows that the inequality is
strict, as claimed.

In summary, we know that $k$ and $h$ agree on the sign of the relativistic
mass of $b$, and because $L$ is norm preserving, we know that
\Norm{\fvp_k(b)} = \Norm{\fvp_h(b)}.  For the relativistic case we have
$\Norm{\fvp_k(b)} = \NormM{\mathsf{m}_k(b) \cdot (1, \vel_k(b))} = \Abs{\mathsf{m}_k(b)}
\cdot \sqrt{ \Abs{1 - v_k^2(b)} }$, and likewise for $h$.  Because
$\mathsf{m}_k(b)$ and $\mathsf{m}_h(b)$ have the same sign, we can remove the relevant
modulus signs, and conclude that \[
  \mathsf{m}_k(b) \cdot \sqrt{ \Abs{1 - v_k^2(b)} } = \mathsf{m}_h(b) \cdot \sqrt{ \Abs{1
- v_h^2(b)} }.
 \] In the Newtonian case, we have $\Norm{\fvp_k(b)} = \NormT{
(\mathsf{m}_k(b), \mathsf{m}_k(b) \cdot \vel_k(b))} = \Abs{\mathsf{m}_k(b)}$, and hence \[
  \mathsf{m}_k(b) = \mathsf{m}_h(b) \] as claimed.
\end{proof}

\section{Concluding remarks}
\label{sec:conclusion}

We began this paper by asking why the masses and momenta of faster-than-light particles decrease with their velocities. In the sense of \cite{Sz11}, Theorem \ref{mainthm} answers this question by explaining that the axioms described in \ax{DYN}+\ax{AxPh} leave no other options open to us. The fact that our axiom system is so sparse, and the axioms themselves so elementary, makes this answer dif\/f\/icult to challenge, and highlights the advantages of the axiomatic approach. If experimental evidence one day shows our conclusions to be physically invalid, this will necessarily point to the invalidity also of one of our axioms, thereby identifying prof\/itable areas for future investigation.

Alternatively, one can see the results presented here as a comment on our underlying axiom system~-- the axioms are strong enough to allow familiar and meaningful results to be derived for slow particles, yet generous enough not to disallow the existence of FTL particles for which reference frame transformations preserve norms as they do for slow particles\footnote{We are grateful to the anonymous referee for this observation.}.

\appendix

\section{Rigorous proof of Lemma~\ref{lem:special-particle}}

Bolzano's theorem does not hold in general in arbitrary Euclidean ordered f\/ields, but it holds for quadratic functions by item (i) of Lemma~\ref{bolzano}, and as we shall see, this is all we require.

Recall that $r$ is a {\bf root} of  a function $f:\Q\rightarrow\Q$ if{}f $f(r)=0$, and that
$f:\Q\rightarrow \Q$ is a~{\bf quadratic function} if{}f there are $a,b,c\in\Q$ such that $f(x)=ax^2+bx+c$ for every $x\in\Q$.

\begin{lemma}
\label{bolzano}
Assume \ax{AxEField}. Then $(i)$ and $(ii)$ below hold.
\begin{itemize}\itemsep=0pt
\item[$(i)$]
Let $f$ be a quadratic function. Assume $p,q\in\Q$ are such that
$p<q$, $f(p)>0$ and $f(q)<0$. Then $f$ has a root between $p$ and $q$,
i.e.\ there is $r\in\Q$ such that $f(r)=0$ and $p<r<q$.
\item[$(ii)$] Let $f:\Q\rightarrow\Q$ be a function such that there are
quadratic functions $f_1$,  $f_2$ and $s\in\Q$
\begin{gather*}
f(x)= \begin{cases}
          f_1(x) & \text{ if $x\leq s$},\\
          f_2(x) & \text{ if $x\geq s$}.
\end{cases}
\end{gather*}
Assume $p,q\in\Q$ are such that $p<q$ and $f(p)>0$ and $f(q)<0$. Then $f$
has a root between $p$ and $q$.
\end{itemize}
\end{lemma}

\begin{proof}
To prove item (i), assume $f(x)=ax^2+bx+c$ for every $x\in\Q$
and $f(p)>0$, $f(q)<0$ and $p<q$.
Without loss of generality we can assume that $a>0$.
Then, for every $x\in\Q$,
\begin{gather*}
f(x)=a\left(x+\frac{b}{2a}\right)^2-\frac{b^2-4ac}{4a}.
\end{gather*}
If $b^2-4ac\leq 0$, then $f(x)\geq 0$ for every $x\in\Q$.
Thus the discriminant $b^2-4ac$ is positive since $f(q)<0$.
Then, by the same methods as for the f\/ield of reals, one can prove that
$f$ has
exactly two
roots $x_1,x_2\in\Q$ such that $x_1<x_2$ and
\begin{gather*}
f(x)=a(x-x_1)(x-x_2) \quad \text{for every $x\in\Q$.}
\end{gather*}
Then $f(x)\leq 0$ if $x\in[x_1,x_2]$ and
$f(x)>0$ if $x<x_1$ or $x>x_2$. By this, we conclude that
$x_1\in[p,q]$, and this completes the proof of~(i).

To prove item (ii) assume $f$, $f_1$, $f_2$, $s$, $p$, $q$ satisfy the assumptions.
Let us note that $f_1(s)=f_2(s)$.

First assume that $s\leq p$. Then $f$ coincides with $f_2$ on $[p,q]$,
i.e.\ $f(x)=f_2(x)$ for every $x\in[p,q]$. Then $f_2$ has a root between
$p$ and $q$ by item (i) of the lemma. Hence $f$ has a root between $p$
and $q$.

Now assume that $s\geq q$. Then $f$ coincides with $f_1$ on $[p,q]$.
Then by item (i) of the lemma we conclude that $f$ has a root between
$p$ and $q$.

Finally assume that $p<s<q$. Then $f$ coincides with $f_1$ on $[p,s]$
and $f$ coincides with $f_2$ on $[s,q]$. If $f(s)<0$, then
$f_1$ has a root between $p$ and $s$, and if $f(s)>0$
then $f_2$ has a root between $s$ and $q$ by item (i) of the lemma.
Therefore, $f$ has a root between $p$ and $q$.
\end{proof}

\begin{proof}[Proof of Lemma~\ref{lem:special-particle}]
Assume $k$, $h$ satisfy the assumptions. Let $L$ be the linear part of
the world-view transformation $\w_{kh}$. In the proof we will need the
following def\/inition.
Let $\vx\in\Q^n$ be such that $x_1\neq 0$.
Then  the \textbf{speed} $v(\vx)$ of vector $\vx$ and $\li(\vx)$ determined
by vector $\vx$ are def\/ined as
 \begin{gather*}
v(\vx)\leteq   \frac{\sqrt{x_2^2+\dots+x_{n}^2}}{|x_1|},\qquad \text{and}\qquad
\li(\vx)\leteq   \setopen q\cdot \vx\setmid q\in\Q\setclose.
\end{gather*}

For every $\bu,\bv\in\Q^3$, let
\begin{gather*}
\bu\prec\bv\ \defiff\ \bu\neq \bv\land \bu\preceq\bv.
\end{gather*}

\noindent
{\it Claim.} Assume $v_k(h)>0$. Then there is $\vu\in\Q^3$ such that
\begin{gather}
\label{lem4-e11}
\vu\prec\vel_k(h)\qquad \text{and}\qquad v(1,\vu)=v\big(L(1,\vu)\big),
\end{gather}
see Fig.~\ref{22d}.
\begin{figure}[t]\centering
\includegraphics{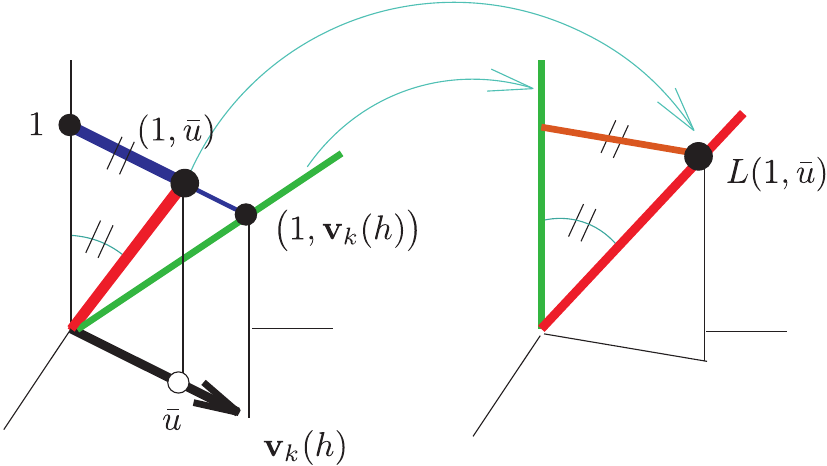}
\caption{Illustration for the claim that \eqref{lem4-e11} can be satisf\/ied.}
\label{22d}
\end{figure}

We will prove the claim at the end of the proof. We note that in the f\/ield
of reals, by continuity and Bolzano's theorem, it is easy to see that the claim holds.

If $v_k(h)=0$, then by \ax{AxThEx^-} there is an inertial particle $b$ such that $v_k(b)=0$ and $\mathsf{m}_k(b)\neq 0$. Then $v_h(b)=0$ and $b$ has the desired properties. Suppose instead, therefore, that $v_k(h)>0$. Then $v_h(k)>0$. Let $\vu\in\Q^3$ be such that \eqref{lem4-e11}
holds. Then, by \ax{AxThEx^-}, there is an inertial particle $b$ such that
$\vel_k(b)=\vu$ and $\mathsf{m}_k(b)\neq 0$. Then $\wl_k(b)$ is parallel with  $\li(1,\vu)$
and $\wl_h(b)$ is parallel with $\li \big(L(1,\vu)\big)$. Thus
$v_k(b)=v(1,\vu)$ and $v_h(b)=v\big(L(1,\vu)\big)$. Thus
$v_k(b)=v_h(b)$ and $\vel_k(b)\prec \vel_k(h)$.

Now we turn to prove the claim. Let $K:\Q^2\rightarrow\Q^4$
and $H:\Q^2\rightarrow\Q^4$  be the linear embeddings such that
\begin{gather*}
K(1,0)=(1,0,0,0)\qquad \text{and}\qquad
K\big(0,v_k(h)\big)=\big(0,\vel_k(h)\big),\qquad \text{and}\\
H(1,0)=(1,0,0,0)\qquad \text{and}\qquad
H\big(0,v_h(k)\big)=\big(0,\vel_h(k)\big).
\end{gather*}
Both $K$ and $H$ preserve
the Euclidean distance and take the time-axis $\li(1,0)$
of $\Q^2$ to the \EMPH{time axis} $\taxis \leteq \{ \vx \in \Q^4 \setmid x_2 =
x_3 = x_4 = 0 \}$ of $\Q^4$. Furthermore,
 for every $(t,x)\in\Q^2$,
\begin{gather}
\label{lem4-uj1}
v(t,x)=v\big(K(t,x)\big)=v\big(H(t,x)\big).
\end{gather}
Let
\begin{gather}
\label{lem4-uj6}
T\leteq H^{-1}\circ L\circ K.
\end{gather}

By \ax{BA}, the world-view transformation
$\w_{kh}$ takes $\taxis$ and $\wl_k(h)$ to $\wl_h(k)$ and $\taxis$,
respectively. Moreover, $\li\big(1,\vel_k(h)\big)$ and $\li\big(1,\vel_h(k)\big)$ are
parallel with $\wl_k(h)$ and $\wl_h(k)$, respectively. Therefore,
\begin{gather*}
  \text{$L$ takes $\taxis$ to
$\li\big(1,\vel_h(k)\big)$ and takes
$\li\big(1,\vel_k(h)\big)$ to $\taxis$}, \\
  \text{$K$ takes $\li(1,0)$ to $\taxis$ and
takes $\li\big(1,v_k(h)\big)$ to
                  $\li\big(1,\vel_k(h)\big)$},\\
  \text{$H$ takes $\li(1,0)$ to $\taxis$, and takes $\li\big(1,v_h(k)\big)$ to
                  $\li\big(1,\vel_h(k)\big)$}.
\end{gather*}
Therefore,
\begin{gather}
  \text{$T$ takes $\li(1,0)$ to $\li\big(1,v_h(k)\big)$},\nonumber\\
  \text{$T$ takes $\li\big(1,v_k(h)\big)$ to $\li(1,0)$},\label{lem4-kell}
\end{gather}
and $T:\Q^2\rightarrow\Q^2$ is a linear transformation.

 Next we prove that
\begin{gather}
\label{lem4-uj3}
\exists\, u\  \big[0<u<v_k(h)\ \land\ v(1,u)=v\big(T(1,u)\big)\big].
\end{gather}
Let $a,b,c, d\in\Q$ be such
that $T(t,x)=(at+bx, ct+dx)$ for every $t,x\in\Q$. Such $a$, $b$, $c$, $d$ exist
and $ad-bc\neq 0$ since $T$ is a linear transformation.
By~\eqref{lem4-kell}, $T(1,v_k(h))_2=0$. Thus $c+d\cdot v_k(h)=0$.
Therefore,
$c=-d\cdot v_k(h)$ and, for every $x\in\Q$,
 $T(1,x)=(a+bx, -d\cdot v_k(h)+dx)$
and
\begin{gather*}
v\big(T(1,x)\big)=\frac{|d|\cdot \big(v_k(h)-x\big)}{|a+bx|}\qquad \text{if $a+bx\neq
0$ and $0\leq x\leq v_k(h)$}.
\end{gather*}
By $c=-d\cdot v_k(h)$ and $ad-bc\neq 0$ we have that
\begin{gather}
\label{lem4-e3}
d\neq 0\qquad \text{and}\qquad a+b\cdot v_k(b)\neq 0.
\end{gather}

Clearly, $v(1,x)=x$ for every $x\geq 0$.
Thus, to prove \eqref{lem4-uj3},
we have to prove that
\begin{gather}
\label{lem4-e22}
\exists\, u\ \left[ a+bu\neq 0\ \land\ 0<u<v_k(h)\ \land \
\frac{|d|\cdot \big(v_k(h)-u\big)}{|a+bu|}=u\right].
\end{gather}
Let us def\/ine a function $g:\Q\rightarrow\Q$ as follows:
\begin{gather}
\label{lem4-masodfoku}
g(x)\leteq |d|\cdot\big(v_k(h)-x\big)- |a+bx|\cdot x.
\end{gather}
If $u$ is a root of $g$, then by~\eqref{lem4-e3},
$a+bu\neq 0$.  Thus, we can prove~\eqref{lem4-e22} by showing that function
$g$ has a root between $0$ and $v_k(h)$.

Clearly,  $g(0)=|d|\cdot v_k(h)$ and
$g\big(v_k(h)\big)=-|a+b\cdot v_k(h)|\cdot v_k(h)$ by
\eqref{lem4-masodfoku}. Thus, by \eqref{lem4-e3},
\begin{gather}
\label{lem4-bolz}
g(0)>0 \qquad \text{and}\qquad g\big(v_k(h)\big)<0.
\end{gather}
By Bolzano's theorem and~\eqref{lem4-bolz}, it is easy to prove that $g$ has
a root between $0$ and $v_k(h)$ if we assume that our f\/ield is the f\/ield
of real numbers. To prove this
for arbitrary Euclidean ordered f\/ields, let $g_0$, $g_1$ and $g_2$
be the following quadratic functions:
\begin{gather*}
g_0(x)   \leteq  |d|\cdot\big(v_k(h)-x\big)- |a|\cdot x, \\
g_1(x)   \leteq |d|\cdot\big(v_k(h)-x\big)- (a+bx)\cdot x, \qquad \text{and}\\
g_2(x)   \leteq  |d|\cdot\big(v_k(h)-x\big)+ (a+bx)\cdot x.
\end{gather*}

Assume that $b=0$. Then, $g(x)=g_0(x)$ for every $x$.  By
Lemma~\ref{bolzano}(i) and \eqref{lem4-bolz}, we conclude that
$g$ has a root between $0$ and $v_k(h)$.

Now, assume $b\neq 0$. Then $i$, $j$ can be chosen such that $\setopen
i,j\setclose=\setopen 1,2\setclose$ and
\begin{gather*}
g(x) =  \begin{cases} g_i(x) & \text{if $x\leq -a/b$,}\\
                               g_j(x) & \text{if $x\geq -a/b$}.
\end{cases}
\end{gather*}
Now, by Lemma~\ref{bolzano}(ii) and~\eqref{lem4-bolz},
$g$ has a root between $0$ and $v_k(h)$.

We have proved that \eqref{lem4-uj3} holds.

Let $u$ be such that
\begin{gather}
\label{lem4-ezaz}
0<u<v_k(h)\ \land\ v(1,u)=v\big(T(1,u)\big).
\end{gather}
Let $\vu\in\Q^{3}$ be such that $K(1,u)=(1,\vu)$. Then $\vu\prec \vel_k(h)$
by def\/inition of $K$ and \eqref{lem4-ezaz}. Furthermore, by \eqref{lem4-uj1},
\eqref{lem4-uj6} and \eqref{lem4-ezaz}, we have that
$v(1,\vu)=v\big(K(1,u)\big)=v(1,u)=v\big(T(1,u)\big)=
v\big(H^{-1}\circ L\circ K(1,u)\big)=v\big(H^{-1}\circ L(1,\vu)\big)=
v\big(L(1,\vu)\big)$. Thus $v(1,\vu)=v\big(L(1,\vu)\big)$. Thus the
claim holds, and this completes the proof of the lemma.
\end{proof}

\subsection*{Acknowledgements}
This research is supported under the Royal Society International
Exchanges Scheme (ref.\ \linebreak  IE110369) and by the Hungarian Scientif\/ic Research
Fund for basic research grants No.~T81188 and No.~PD84093, as well as by a
Bolyai grant for J.X.~Madar\'asz.

\pdfbookmark[1]{References}{ref}
\LastPageEnding

\end{document}